\documentclass{ws-procs10x7}
\usepackage{balance}

\newcolumntype{d}[1]{D{.}{.}{#1}}

\def\Journal#1#2#3#4{{\it #1} {\bf #2}, #3 (#4)}

\makeindex
\begin{document}

\title{Mini review on saturation and recent developments}

\author{CYRILLE MARQUET}

\address{Service de Physique Th\'eorique, CEA/Saclay, 91191 Gif-sur-Yvette Cedex, France\\URA 2306, unit{\'e} de recherche associ{\'e}e au CNRS
\\E-mail: marquet@spht.saclay.cea.fr}


\twocolumn[\maketitle\abstract{I discuss the saturation regime of QCD: the
weak-coupling regime that describes large parton densities inside hadrons and the resulting high-energy behavior of scattering amplitudes. I briefly review past successes in the context of deep inelastic scattering at HERA and forward particle production at RHIC, and I present some recent progresses that significantly improved our understanding of high-energy scattering in QCD.}
]


The QCD description of hadrons in terms of quarks and gluons depends on the processes considered and on what part of the hadron wavefunction they are sensitive to. Consider a hadron moving at nearly the speed of light along the light cone direction $x^+,$ with momentum $P^+.$ Depending on their transverse 
momentum $k_T$ and longitudinal momentum $xP^+,$ the partons inside the hadron behave differently, reflecting the different regimes of the hadron wavefunction. 

When probing the (non-perturbative) soft part of the wavefunction, corresponding to partons with transverse momenta of the order of $\Lambda_{QCD}\!\sim\!200\ \mbox{MeV},$ the hadron looks like bound state of strongly interacting partons. When probing the hard part of the wavefunction, corresponding to partons with $k_T\!\gg\!\Lambda_{QCD}$ and $x\!\lesssim\!1,$ the hadron looks like a dilute system of weakly interacting partons.

The present work deals with the small$-x$ part of the wavefunction and with the so-called saturation regime of QCD. When probing partons that feature
$k_T\!\gg\!\Lambda_{QCD}$ and $x\!\ll\!1,$ the effective coupling constant
$\alpha_s\log(1/x)$ is large, and the hadron looks like a dense system of weakly interacting partons, mainly gluons (called small$-x$ gluons).

The larger $k_T$ is, the smallest $x$ needs to be to enter the saturation regime. As pictured in Fig.1, this means that the separation between the dense and dilute regimes is characterized by a momentum scale $Q_s(x),$ called the saturation scale, which increases as $x$ decreases. The scattering of dilute partons (with $k_T\!\gg\!Q_s(x)$) is described in the leading-twist approximation in which they scatter incoherently. By contrast, when the parton density is large $(k_T\!\sim\!Q_s(x)),$ partons scatter collectively. The saturation regime of QCD is the perturbative regime that describes this collective behavior.

To be sensitive to the small$-x$ physics, high energies are needed. Hence before the mid 90's, saturation was not relevant for any experimental measurements, contrary to soft QCD (for instance in hadron-hadron elastic scattering) or hard QCD (for instance in top quark production). But with the start of HERA and then RHIC, there has been a recent gain of interest for saturation physics as observables sensitive to the small$-x$ part of the hadron wavefunction could be measured.


One of the most studied process is deep inelastic scattering (DIS): a photon with virtuality $Q^2$ probes the hadron wavefunction and observables are sensitive to values of $x$ as small as $x_{Bj}\!\simeq\!Q^2/s.$ At high energy
$s\!\gg\!Q^2,$ both inclusive, diffractive and exclusive processes are probing the saturation regime.

\begin{figure}[t]
\centerline{\psfig{file=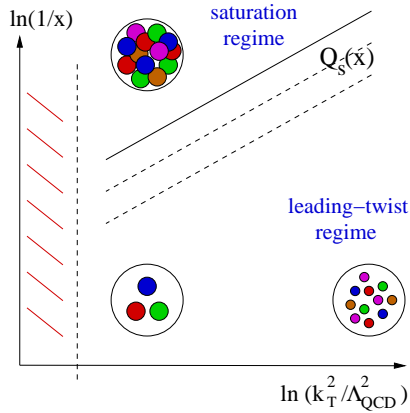,width=5.3cm}}
\caption{Diagram in the $(k_T,x)$ plane picturing the hadron in the different weakly-coupled regimes. The saturation line separates the dilute (leading-twist) regime from the dense (saturation) regime.}
\end{figure}

In DIS, the hadronic scattering can be viewed as that of a colorless $q\bar q$ pair (or dipole) of transverse size $r\!\sim\!1/Q$ off the hadron. The evolution of the dipole scattering amplitude $T(r,x)$ with decreasing $x$ is described by the B-JIMWLK equation~\cite{jimwlk,cgc}, established in the leading $\ln(1/x)$ approximation, and by its mean-field version the BK equation~\cite{bk}.

What the dipole sees is what is pictured in Fig.1 with $k_T\!\rightarrow\!1/r.$ In the leading-twist regime $T\ll1,$ and in the saturation regime $T=1.$ From the BK equation and its equivalence with the FKPP equation, well known in statistical physics, it was shown that the growth of the dipole amplitude towards the saturation regime occurs in a particular way~\cite{gstheo}. When approaching the saturation regime, instead of being a function of a priori the two variables $r$ and $x,$ the dipole scattering amplitude is actually a function of the single variable $rQ_s(x)$ up to inverse dipole sizes significantly larger than the saturation scale $Q_s(x).$

This means that in the diagram shown Fig.1, lines parallel to the saturation line are lines of constant densities, along which $T$ is constant. This property which extends towards small values of $r$ (or large values of $Q^2$) manifest itself in the data. Indeed, this implies the geometric scaling of the total cross-section at small $x:$
\begin{equation}
\sigma^{\gamma^*p\rightarrow X}_{tot}(x,Q^2)=
\sigma^{\gamma^*p\rightarrow X}_{tot}(\tau\!=\!Q^2/Q_s^2(x))\ .
\end{equation}
As shown in Fig.2, this has been confirmed by experimental data~\cite{gsincl} with $Q_s(x)=(x/x_0)^{-\lambda/2}\ \mbox{GeV}$ and the parameters
$\lambda\!=\!0.288$ and $x_0\!=\!3.04\ 10^{-4}$ taken from~\cite{gbw}.

\begin{figure}[t]
\centerline{\psfig{file=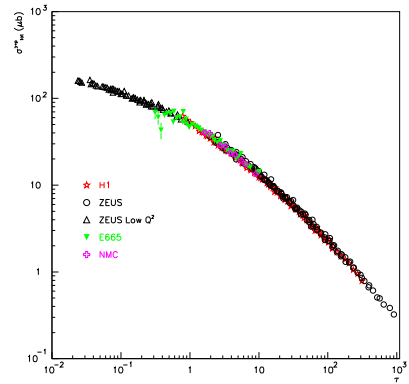,width=5.3cm}}
\caption{The cross section $\sigma^{\gamma^*p\rightarrow X}_{tot}$ as a 
function of $\tau\!=\!Q^2/Q_s^2(x)$ for $x\!<\!0.01.$}
\end{figure}

The scaling law $T(r,x)=T(rQ_s(x))$ also implies the geometric scaling of the diffractive cross section at small $x_{\mathbb P}$ and fixed $\beta,$ with the same saturation scale:
\begin{eqnarray}
\sigma^{\gamma^*p\rightarrow Xp}_{diff}(\beta,x_{\mathbb P},Q^2)=
\hspace{2.5cm}\nonumber\\
\sigma^{\gamma^*p\rightarrow Xp}_{diff}
(\beta,\tau_d\!=\!Q^2/Q_s^2(x_{\mathbb P}))\ .
\end{eqnarray}
This is in agreement~\cite{gsdiff} with the data from HERA (see Fig.3). More quantitative analysis have been carried out: saturation
models~\cite{gbw,bgbk,iim} fit well $\sigma^{\gamma^*p\rightarrow X}_{tot}$ data and give accurate predictions for $\sigma^{\gamma^*p\rightarrow Xp}_{diff},$ and more exclusive processes like deeply virtual Compton scattering and vector meson production~\cite{ismd}.

\begin{figure}[t]
\centerline{\psfig{file=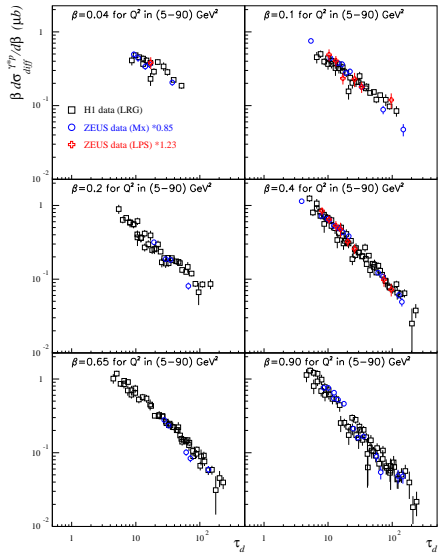,width=5.8cm}}
\caption{The cross section
$\beta\ d\sigma^{\gamma^*p\rightarrow Xp}_{diff}/d\beta$ as a function of
$\tau_d\!=\!Q^2/Q_s^2(x_{\mathbb P})$ in bins of $\beta$ for
$x_{\mathbb P}\!<\!0.01.$}
\end{figure}

More studies contributed to the success of saturation in the context of forward particle production in hadron-hadron collisions. When producing particles (jets, pions, dileptons, heavy flavors) with transverse momentum $p_T$ and rapidity 
$y,$ the typical values of $x$ probed in the hadrons wavefunctions are $x_{1,2}\!\sim\!p_T\ e^{\pm y}/\sqrt{s}.$ Therefore particle production at forward rapidities and high energies involves high values of $x$ for one hadron and small values of $x$ for the other. To describe such a process, the dipole picture can also be used~\cite{moi}, allowing many investigations~\cite{jamyu}.

Among other things, saturation predicted the high-rapidity suppression of the nuclear modification factor in d-Au collisions, which was observed at RHIC. Other predictions, such as the suppression of back-to-back correlations, are in qualitative agreement with the data.


Recently, some limitations of the B-JIMWLK equation were pointed 
out~\cite{trig1,trig2}. This triggered a series of papers that aimed at completing the B-JIMWLK equation by including Pomeron loops~\cite{ploop1,ploop2} in the high-energy evolution. These studies improved our understanding of different aspects of high-energy scattering. For instance, they allowed the constructions of an effective action~\cite{heea}, a generalized dipole 
model~\cite{gdm} and a reggeon field theory~\cite{rft}.

In the following, we discuss the phenomenological consequences of Pomeron loops.
Within the large$-N_c$ limit, the dipole amplitude $T(r,x)$ can be obtained from a Langevin equation~\cite{ploop1} that features a noise term, similar to the stochastic FKPP equation~\cite{trig2}. Its solution is an event-by-event dipole scattering amplitude $\tilde{T}(rQ_s(x))$ characterized by a saturation scale $Q_s$ which is a stochastic variable. Using results derived in the weak-noise~\cite{wnoise} and strong-noise~\cite{snoise} limits, it was shown that
$\ln(Q_s/Q_0)$ (with $Q_0$ a reference scale) is distributed according to a Gaussian probability law~\cite{probdis}. Denoting $Y\!=\!\ln(1/x),$ the average value is $\ln(\bar{Q}_s^2/Q_0^2)\!=\!\lambda Y$ and the variance is
$\sigma^2\!=\!DY.$

One obtains the physical amplitude $T$ after averaging
$\tilde{T}$ over the different realizations of the noise. The dispersion coefficient $D$ allows to distinguish between two energy regimes: the geometric scaling regime ($DY\!\ll\!1$) and diffusive scaling regime ($DY\!\gg\!1$). In formulae, the averaged amplitude features the following scaling behaviors:
\begin{eqnarray}
T(r,Y)\!&\stackrel{Y\ll1/D}{=}&\!T_{gs}\left(r\bar{Q}_s(Y)\right)\ ,\label{gs}\\
T(r,Y)\!&\stackrel{Y\gg1/D}{=}&\!T_{ds}\left(\ln(r\bar{Q}_s(Y))/\sqrt{DY}\right)\ .\hspace{0.3cm}\label{ds}
\end{eqnarray}
In the saturation region $r\bar{Q}_s\!>\!1,$ $T(r,Y)\!=\!1.$ As the dipole size $r$ decreases, $T(r,Y)$ decreases towards the dilute regime, following the scaling laws (\ref{gs}) or (\ref{ds}), depending on the value of $DY.$

In other words, Pomeron loops turn the geometric scaling regime into an intermediate energy regime, in which the dispersion of the events is negligible. It predicts the emergence of a new scaling law at very high energies, when the event dispersion is important: diffusive scaling~\cite{diffscal}. This is summarized in Fig.4 (with $r\!\rightarrow\!1/Q,$ Fig.4 applies also to DIS). A key feature of the diffusive scaling regime is that, up to inverse dipole sizes much larger than the average saturation scale, cross-sections are dominated by events which feature the hardest fluctuation of the saturation scale. In average the scattering is weak yet saturation is the relevant physics. 

While it seems that HERA and RHIC are probing the geometric scaling regime, this has important implications for the LHC. In the context of forward particle production~\cite{dslhc}, the energy there might be high enough to reach the diffusive scaling regime. However, our poor theoretical knowledge of the dispersion coefficient $D$ prevents more quantitative statements at the moment.

\begin{figure}[t]
\centerline{\psfig{file=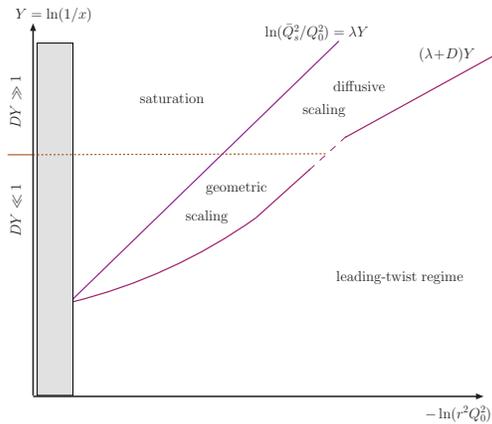,width=6.5cm}}
\caption{A phase diagram for the high-energy limit of DIS in QCD. Shown are the average saturation line and the boundaries of the scaling regions at large values of $\ln(1/r^2).$ With increasing $Y,$ there is a gradual transition from geometric scaling at intermediate energies to diffusive scaling at very high energies.}
\end{figure}

\balance

\end{document}